\documentclass{article}
\usepackage[utf8]{inputenc}
\usepackage{authblk}

\title{Private Federated Learning in Gboard}
\author{Yuanbo Zhang, Daniel Ramage, Zheng Xu, Yanxiang Zhang, Shumin Zhai, Peter Kairouz}
\affil{Google}
\date{June, 2023}

\usepackage{graphicx}
\usepackage{parskip}
\usepackage[margin=1in]{geometry}
\usepackage[
backend=biber,
sorting=none,
maxbibnames=99,
]{biblatex}
\bibliography{references}

\begin{document}

\maketitle

\section{Introduction}

Gboard relies on machine learning (ML) innovations to improve users’ input experience with the product, which benefits from statistics and models derived from many users’ typing data. To meet these needs, Gboard invests in privacy technologies that allow the data to be processed locally on device, to be aggregated as early as possible, and to have strong anonymization and differential privacy where possible.

Gboard has been one of the earliest adopters of federated learning (FL) \cite{flsysdesign} at Google, which enabled mobile devices to collaboratively train models while keeping raw training data such as typing content on each user's device. In the past few years, various language models and correction models have been built with FL to optimize Gboard input efficiency and quality by providing more accurate suggestions, auto corrections, smart compose prediction and next word predictions \cite{kbnwp}.

Gboard has deployed FL models using a variety of privacy enhancing technologies with an eye toward a fully privacy preserving ML technology stack. This white paper describes recent advances that Gboard has made in bringing federated learning, secure aggregation, and differential privacy techniques to Gboard’s users through a multi-year effort combining research and product integrations, and looking forward, we observe how technologies such as trusted execution environments (TEEs) may play a larger role in supporting features powered by next-generation generative AI models.

\section{Privacy Principles and Practices}

Privacy is multi-faceted. Following the broader discussion of privacy for FL introduced in \cite{flprivacy}, we focus on several specific aspects of privacy:

\begin{enumerate}
\item \textbf{Transparency and user control:} users can be aware of what data is used, what purpose it is used for and how it is processed, and have full control on whether to enable the collection and use of their data.
\item \textbf{Data minimization:} data is only collected focusing on specific computation needs, aggregated as early as possible and discarded as soon as possible, with access limited at all data processing stages.
\item \textbf{Data anonymization:} the final released output of the computation does not reveal anything unique to an individual.
\end{enumerate}

In Gboard, whether FL and other types of machine learning are enabled is in users’ control, and can be easily configured through Gboard settings \cite{gboardprivatesetting}.

\begin{figure}[h]
    \centering
    \includegraphics[width=0.8\textwidth]{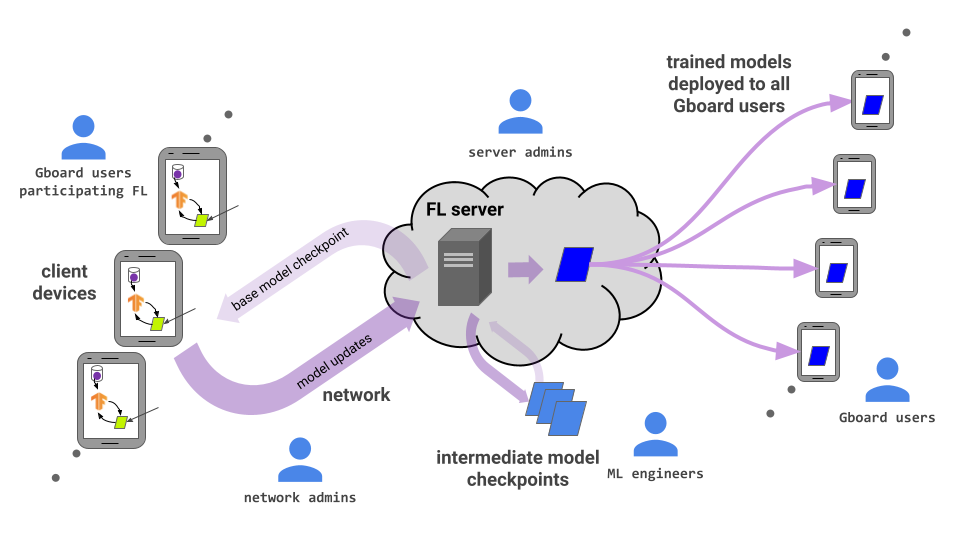}
    \caption{Gboard FL system overview}
    \label{fig:system_overview}
\end{figure}

Data minimization and data anonymization are achieved by a combination of technologies and policies at different data processing stages. As illustrated in the Figure \ref{fig:system_overview},  federated learning makes progress in discrete training rounds. In each round, a set of currently available devices (from hundreds to thousands) are sampled from the training population, download a base model checkpoint (an initial model or the output of the previous round from the FL server), run stochastic gradient descent (SGD) for optimization on their own data locally, and upload model updates to server for immediate aggregation. The server aggregates model updates by averaging them, further processes the aggregate if needed, and produces a new checkpoint as an intermediate output. After sufficient rounds of updates, a final model checkpoint may be further processed (e.g. compressed or quantized), evaluated for quality, and rolled out for use in inference for Gboard’s users.

Google’s production FL service is a multi-tenant service operated by a service team separate from Gboard. Google’s FL service does not log or grant access to unaggregated model updates. Model updates are immediately aggregated within the isolated service’s boundaries as enforced by Access Control Lists (ACLs), Multi-Party Authorization \cite{mpa}, production job role accounts, code reviews, etc. The isolation boundary is hardened with complementary technologies including distributed differential privacy (DDP) \cite{ddpsecagg, skellamdp, tffddpsecagg} and Secure Aggregation (SecAgg) \cite{practicalsecagg} when so configured. (Gboard’s use of DDP and SecAgg is discussed below.) With these stronger protections and under a correctly implemented key exchange and aggregation protocol (i.e. an honest-but-curious server), even the aggregation service itself has no access to unaggregated updates. For data transmission security, Gboard makes gRPC or HTTPS requests to the FL system from within the Federated Compute Client Open Source (OSS) code \cite{httpfedprotocol}, \cite{grpcfedprotocol}. Network requests include an API key for endpoint authorization not specific to a device and includes an attestation token to protect against botnets via an internal service similar to the Play Integrity API \cite{playintegrity}. The attestation token used by FL contains no device identifiers, and is not stored, logged, or analyzed by the FL service, which uses the attestation token only to verify client integrity with the attestation backend. The API key, attestation token, and connection metadata including source IP, are all discarded by FL service before aggregation. The service team makes use of telemetry logs (without persistent device identifiers) in order to properly operate the service.

Figure~\ref{fig:system_overview} also illustrates the access points in the system. ML engineers subject to use-case specific ACLs may access model checkpoints after aggregation of each training round. Final models may be accessed by many more parties after public deployment. We therefore place the strongest data anonymization technologies such as formal user-level differential privacy guarantees to models going back to user devices. Gboard uses DP-FTRL \cite{dpftrl} to bound how much the model parameters can memorize content unique to a particular individual. Case-by-case privacy reviews are required for all new uses of FL in Gboard, with code review used to guard against mistakes and to enforce required minimum aggregation sizes, DP and/or SecAgg requirements if appropriate, etc.

\section{Towards Differentially Private FL for Gboard Language Models}

Language models (LM) play an essential role in Gboard suggestion, correction and prediction functions. Ideally, LMs trained on user data should capture common patterns typed by many users, but should not encode content unique to a particular individual - i.e. LM parameters should be computed with appropriate \textit{data anonymization}. The relevant notion of privacy would be differential privacy,  i.e. that the model parameters learned during training are not statistically distinct given a small change to the training data. This can be quantified as a "small” $\rho$ value of zero-Concentrated-Differential-Privacy (zCDP) or $\varepsilon$ of $(\varepsilon, \delta)$-Differential Privacy (DP). A common use of DP is \textit{example-level}, where adding or removing a single training example changes model parameters in a provably minimal way. However, as users can contribute multiple examples to the training dataset, example-level DP may not be strong enough to ensure the users’ data isn't memorized. Instead, we choose to use \textit{user-level} DP, i.e. that the model parameters learned during training are not statistically distinct even if we add/remove all the training examples from any one particular Gboard user.

In Feb 2022 we deployed the DP-FTRL (Differentially Private Follow-the-Regularized-Leader)  algorithm trained LM for next word prediction for Spanish language Gboard users, which for the first time offered a formal DP guarantee in production (with zCDP $\rho=0.81$) \cite{dpftrlblog}. Since then we have been working on scaling DP-FTRL to learn LMs with larger parameter sizes for different features and on different user populations of various sizes. As of Feb 2023, DP-FTRL has been widely adopted in LM training to enable typing features, including neural network based next word prediction, smart compose and decoding suggestions on-the-fly rescore for British English, French, German, India English, Italian, Norwegian, Portuguese, Russian, Spanish and US English language Gboard users. In parallel, we are also refreshing existing FL LMs for all other languages. 

We continuously enhance the performance of DP algorithms, practices and settings over time and look for better DP guarantees without utility loss (even when comparing with existing high quality models previously trained without DP). Empirically, the $\rho$ value is determined by the level of noise added, bounded device participation through sampling without replacement policy, and total number of training rounds. As a result, it's a challenge to maintain utility (model quality) while keeping strong privacy guarantees, especially when training population size is small. We have employed model pre-training on public C4 corpus based on Common Crawl dataset \cite{c4dataset} combined with training configuration tuning, and achieved almost neutral utility change measured by \textit{prediction picked ratio}: the ratio of picked prediction candidates among the shown NWP predictions, and \textit{prediction accuracy}: the ratio of prediction candidates matching the finally committed words among the NWP model predictions, for a $\rho=1.31$ zCDP model (US English) compared to the previous model trained without DP \cite{gboardlmdpftrl}. Other DP-FTRL trained LMs have been adopting this optimization practice.

Today all new FL tasks training LMs for Gboard production features require DP, with DP-FTRL recommended. We are on track to upgrade all existing FL LMs that have previously been launched without DP to be trained with DP in 2023.

\section{Towards Secure Aggregation and Distributed DP}

While DP-FTRL provides data anonymization after gradient aggregation on the centralized server, data minimization can be strengthened by limiting what data the centralized server sees \cite{flprivacy}, achieved by technologies such as secure aggregation (SecAgg) and distributed differential privacy (DDP).

SecAgg runs a cryptographic secure multi-party computation (MPC) protocol during FL, such that the server can compute the mean update, but cannot see updates of individual devices, which minimizes access to gradient updates of individual devices potentially carrying sensitive information. Since payloads are encrypted, SecAgg also protects network traffic between client and server. SecAgg gives protection guarantees under the honest-but-curious server threat model, that is, the server and a sufficiently large number of the participants behave honestly to execute the protocol, and the key exchange portion of the secure aggregation protocol is correctly implemented. Under that assumption, a server implementing the secure aggregation protocol cannot see the contents of any individual update.

To further protect the server from getting sensitive information in the mean update, DDP extends SecAgg to ensure that model-updates are slightly perturbed before they are sent to the server using the SecAgg protocol, adding another layer of protection, and also leading to a more bandwidth and memory efficient version of SecAgg \cite{ddpsecagg}.

Gboard has been exploring adopting SecAgg/DDP with FL, and we have achieved promising results. Based on research and infrastructure advancements, such as subgraph SecAgg protocol \cite{subgraphsecagg}, which largely reduced communication and computation overheads, Gboard has trained various models with SecAgg, ranging from smaller sized classification models such as key-by-key correction triggering models ($\sim1k$ parameters) and auto correction triggering models ($\sim1.3m$ parameters), to larger sized language models ($\sim3m$ parameters). The key-by-key correction triggering model has been launched in production for English users with neutral quality impact compared with the baseline model without SecAgg. For LM, we are exploring combining with DP-FTRL for production launch, which is continued in Section 5.

\section{Towards Combining Central and Distributed DP}

Central differential privacy with DP-FTRL and distributed differential privacy with SecAgg/DDP protect against different types of threat vectors at different stages of model training, processing and deployment flow. DP-FTRL provides formal guarantees that models that are aggregated on the server side and to be deployed publicly won't memorize training data unique to a particular individual. SecAgg makes sure an honest-but-curious FL server cannot learn individual updates. DDP extends SecAgg to ensure that model-updates are slightly perturbed before they are sent to the server using the SecAgg protocol, adding another layer of protection even against a malicious server. Their combination would allow Gboard to have an effective way to provide more complete end-to-end privacy protection for data minimization and anonymization. Gboard prioritized DP-FTRL for anonymization with formal DP guarantee on content sensitive models such as LMs, given deployed models have a broader attack surface, and there have been stringent security enforcements on the isolated FL server such as ACLs, production job role accounts, code reviews, and so on. We have also been improving SecAgg/DDP settings for training efficiency and testing their combination with DP-FTRL. In June 2023, we launched the first ever LM trained with DP-FTRL + SecAgg/DDP for US English next word prediction task. The model achieved the same utility measured by \textit{prediction picked ratio} and \textit{prediction accuracy} compared the the previously launched model trained with DP-FTRL only, and offered a even better zCDP with $\rho=0.25$ (compared to $\rho=1.31$) by improving practice on pre-training, sampling without replacement policy and so on \cite{gboardlmdpftrl}. We aim to enable their combination for more Gboard typing FL neural models in 2023. 

\section{Future Possibility of External Verifiability of Privacy Claims}

Over time, we aim to make more of Gboard’s privacy, security, and correctness claims externally verifiable, to client devices and more, under a variety of threat models. As richer Trusted Execution Environment (TEE) and trusted runtime binaries \cite{prjoak, fcpfedagg} become available, further possibilities emerge. For example, can SecAgg be hardened in an OSS trusted aggregator runtime, which is verifiably built to prove the claim of honestly running the protocol? It may also open opportunities to design new types of learning or analytics tasks that involve raw data transmission to the trusted aggregator running on TEE, with guarantee that data leaving the trusted boundary is securely aggregated and differentially private.

\section{Conclusion}

We have made various efforts to advance privacy protection practices in Gboard toward stronger privacy guarantees while learning useful signals to better serve our users. This includes DP-FTRL, which offers a formal central DP guarantee for data anonymization, and SecAgg/DDP to further harden the isolation boundary of data access by limiting what data the centralized server sees using cryptographic secure multi-party computation protocol and adding local noise. As they are separately used to learn different Gboard models over time and proven to achieve both good utility and privacy guarantee, we have been working on their combination to provide more complete end-to-end privacy protection for data minimization and anonymization. Recent development of TEE infrastructure and OSS trusted runtime binaries at Google enables the possibility of verifiable privacy claims, which is an open area for us to further explore. We believe these work streams lead towards maximizing values of ML on Gboard user experience, at the same time minimizing potential privacy costs for our users participating in FL training.

\printbibliography
\end{document}